\documentclass[twocolumn,aps,prb,showpacs,preprintnumbers,amsmath,amssymb]{revtex4}
\usepackage{graphicx}
\usepackage{bm}
\usepackage{gensymb}
\usepackage{color}
\begin{document}
\title{Electronic Structure, Phonons and Dielectric Anomaly in Ferromagnetic 
Insulating Double Perovskite La$_{2}$NiMnO$_{6}$}
\author{Hena Das$^{1}$, Umesh V. Waghmare$^{2}$, T. Saha-Dasgupta\footnote{Corresponding Author: tanusri@bose.res.in}$^{,1}$ and 
D. D. Sarma$^{3,4}$}
\affiliation{$^1$ Satyandranath Bose National Centre for Basic Sciences,
Kolkata 700098, India} \affiliation{$^2$  Jawaharlal Nehru Centre for Advanced
Scientific Research,  Jakkur, Bangalore 560064, India}
\affiliation{$^3$ Centre for Advanced Materials, 
Indian Association for the Cultivation of Science, Jadavpur, Kolkata 700032,
India}\affiliation{$^4$ Solid State and Structural Chemistry Unit,
Indian Institute of Science, Bangalore-560012, India}
\date{\today}
\begin{abstract}
Using first-principles density functional calculations, we study the electronic
and magnetic properties of ferromagnetic insulating double-perovskite compound
 La$_2$NiMnO$_6$, which has been reported to exhibit interesting magnetic field
sensitive dielectric anomaly as a function of temperature. Our study reveals
existence of very soft infra-red active phonons that couple strongly with spins
at the Ni and Mn sites through modification of the super-exchange interaction.
We suggest that these modes are the origin for observed dielectric anomaly in 
La$_2$NiMnO$_6$.
\end{abstract}
\maketitle
%%{\it Introduction -}
Double perovskite La$_2$NiMnO$_6$ (LNMO) is an interesting 
compound which is a ferromagnetic (FM) insulator with a Curie temperature 
close to room temperature (Tc $\sim$ 280 K). Recently large 
magnetic field induced changes in dielectric properties have been observed in 
this compound\cite{1}, which makes this compound a promising 
material for potential device applications\cite{2}. 
In spite of great technological importance, the theoretical 
effort for understanding this material is rather limited.
To our knowledge, there exists only one report of linear augmented 
plane wave based basic
electronic structure calculations of LNMO,  
by Matar et al.\cite{3}. 

%%\noindent
In this letter, we carried out first-principles density functional 
 calculations to understand the ferromagnetic insulating behavior in this
 compound as well as to understand the origin of dielectric anomaly that has 
been observed experimentally\cite{1}. We used a combination of two types of 
methods, namely:
(a) muffin-tin orbital based linear muffin-tin orbital\cite{4} and N-th
order muffin-tin orbital (NMTO)\cite{5} methods, and (b) plane wave-based 
methods.   
In the latter, we used ultra-soft pseudopotentials with an energy cutoff of 25 Ry (150 Ry)
on the plane wave basis for wave functions (charge density) and a 6$\times$6$\times$6
mesh of k-points in sampling the Brillouin zone for a phase with unit cell containing
one formula unit and equivalent for other phases.
In particular, the structural optimization and phonon calculations
have been carried out using QUANTUM ESPRESSO\cite{6}
and effective charges and dielectric response have been carried out using
ABINIT\cite{abinit}.
The analysis of hopping interactions by constructing effective 
orbitals, on the other hand, has been carried out within the framework of NMTO.
In our LMTO and NMTO calculations, we have used four different empty spheres 
to achieve the space filling. 
We used a spin polarized generalized gradient approximation (GGA)\cite{PBE} to
 the exchange correlation functional.  

\begin{table}
\caption{Energy-minimized structural parameters of LNMO. Lattice constants have
been kept constant at the experimental values\cite{8_1}.}
\begin{tabular}{cccccccc}
\hline
\hline
\multicolumn{7}{c}{Rhombohedral} \\
%\hline
a(\AA)  &  b(\AA)  &  c(\AA)  & &  x  &  y  &  z  \\
5.474  &  5.474  &  5.474  & La &  0.24980  &  0.24980  &  0.24980 \\
${\alpha}$ &${\beta}$  &${\gamma}$ & Ni&0.0 & 0.0 & 0.0\\
60.671 &60.671 &60.671 & Mn & 0.5 & 0.5 & 0.5\\
       &       &       & O & 0.81403 & 0.67182 & 0.25889 \\ %\hline
\end{tabular}

\begin{tabular}{cccccccc}
%\hline
\multicolumn{8}{c}{ Monoclinic } \\ %\hline
 &a(\AA) & b(\AA)  &c(\AA)& & x & y & z  \\
  &5.467 &5.510 &7.751 & La &0.00838 & 0.03781 &0.24968  \\
  &$\alpha$ &$\beta$  &${\gamma}$& Ni&0.0 & 0.5 & 0.0  \\
  &         &         &          & Mn & 0.5 & 0.0 & 0.0 \\
 &90.000 &90.119 &90.000 & O1 & 0.22344 & 0.20903 & 0.04140  \\
 &   &      &   & O2 & 0.29189 & 0.27756 & 0.45700  \\
 &   &      &   & O3 & 0.42219 & 0.01454 & 0.24243  \\
\hline
\hline
\end{tabular} 
\end{table}

\begin{figure}
%\centering
\includegraphics[width=9cm]{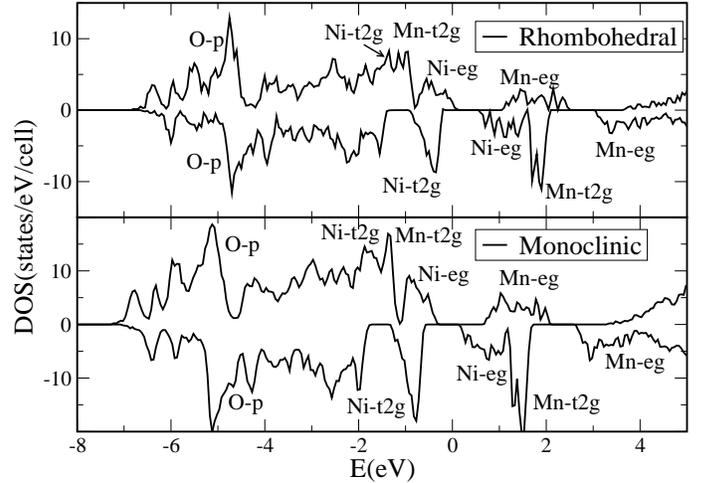}
\caption{\label{fig-1} GGA DOS of LNMO in geometry optimized rhombohedral and
monoclinic phases. Zero of the energy is set at the GGA Fermi energy.}
\end{figure}

%%\noindent     
%%{\it Crystal Structure-} 
%%Double ordered perovskite (A$_2$BB$^{\prime}$O$_6$) possesses a modified 
%%perovskite structure (ABO$_3$) where transition metal based 
%%BO$_6$ and B$^{\prime}$O$_6$ octahedra are alternately arranged along three 
%%cubic or nearly cubic directions
%%and the alkaline earth or lanthanide A atoms occupy body-centered sites of the cubes defined by eight adjacent oxygen octahedra. 
LNMO, having the general structure of a double ordered 
perovskite (A$_2$BB$^{\prime}$O$_6$), is distorted 
from the ideal double perovskite, and 
the amount of distortion changes as the temperature varies. 
The structure of La$_2$NiMnO$_6$ is rhombohedral (R$\bar{3}$) at high
temperature and transforms to monoclinic (P2$_1$/n) at low temperature, 
with these two structures coexisting over a wide temperature range\cite{8,9}.
In view of the fact that the positions of light atoms are often not 
well characterized within the experimental technique,
we have carried out structural optimization of both rhombohedral (RH) and 
monoclinic (MC) phases where the internal degrees of freedom associated with La and 
O atoms have been optimized keeping the lattice parameters fixed at 
experimentally determined values\cite{8}. The relaxed structural 
parameters of the rhombohedral FM state (see Table I) agree well within 
3 \% with experimental ones proving the reliability of our calculation scheme.
The position of oxygen atoms, in particular, the x co-ordinate of O3 oxygen
of the MC phase, however differ noticeably (a deviation of about 
22 $\%$) from the experimental values (compared Table 3 in ref\cite{8}). 
Our results may provide basis to further refinement of the experimental 
structure. Each NiO$_6$ octahedra in the 
rhombohedral phase is tilted with respect to MnO$_6$ octahedra 
giving rise to Ni-O-Mn bond angle of 157$^{\circ}$. The tilting is
further increased by 2$^{\circ}$ in the monoclinic phase.We determined 
the electronic structure of geometry optimized 
FM LNMO, for rhombohedral and monoclinic phases using 
the LMTO\cite{4} basis, as well as 
using the plane wave basis\cite{6}. Both methods resulted in 
similar features in the computed density of states and band structures, and 
insulating solutions for both RH and MC phases. 
The spin resolved partial density of states (DOS) of LNMO in the rhombohedral phase,
calculated using ESPRESSO, is shown in upper panel of Fig.\ref{fig-1}. Below -2 eV the predominant 
contribution is from O-2p states.
The octahedral surrounding of Mn and Ni atoms split the
Mn and Ni d-manifolds into t$_{2g}$ and e$_g$ levels.
In the up spin channel, the Ni-t$_{2g}$ and Ni-e$_g$ levels are 
found in the energy range $\sim$ -2 eV to Fermi energy and show a significant 
mixing with Mn-d states and O-p states. In the down spin channel the 
Ni-t$_{2g}$ 
bands are located between O-p states and Fermi level, while Ni-e$_g$ states  
lye $\sim$ 1.2 eV above the Fermi level. This correspond to the nominal valence of Ni$^{2+}$ (d$^{8}$: t$_{2g}^{6}$e$_{g}^{2}$). 
In the up spin channel the Mn-t$_{2g}$ bands are localized between Ni-t$_{2g}$ and 
Ni-e$_{g}$ bands and are 
filled, while the Mn-e$_g$ bands are separated by a gap 
of $\sim$ 2.5 eV 
from the Mn-t$_{2g}$ bands and remains empty. In the down spin channel, 
both Mn-t$_{2g}$ and Mn-e$_g$ bands are located above Fermi level in the 
energy range 
$\sim$  1.5 eV to 5 eV. This leads to conclusion that the oxidation state of Mn is 
nominally 4+ (d$^{3}$: t$_{2g}^{3}$e$_{g}^{0}$), which agree with the Mn NMR and X-ray absorption spectroscopy 
results\cite{10,11}, though disagree with one of neutron 
diffraction study\cite{9}.
Our spin-polarized LMTO calculations gave a moment of 3.0 $\mu_B$ at the Mn site 
within a muffin-tin(MT) radius of 1.32 \AA, which agree with experimental 
value of 3.0 $\mu_B$\cite{1}. The magnetic moment at the Ni site, for a MT 
radius of 1.52 \AA  is found to
be 1.43 $\mu$B, which is less than the experimentally measured value 
of 1.9 $\mu$B. The residual moment is found to reside at the oxygen sites 
giving rise to the total magnetization of 5.0 $\mu$B in agreement with the experimental 
value\cite{1}.
The spin resolved density of states in the monoclinic 
structure is shown in the lower panel of Fig.\ref{fig-1}. The basic nature of 
DOS is similar to that of rhombohedral phase of LNMO. The occupation of the 
Ni-d states and Mn-d states suggests again the nominal oxidation states of Ni 
and Mn ions to be 2+ and 4+, respectively. The moments are found to be 2.91 $\mu$B 
within a MT radius of 1.38 \AA at the Mn site, 1.35 $\mu$B 
at the Ni site within a MT sphere of radius 1.52 \AA, and $\sim$ 0.10 $\mu$B 
(for 0.95 \AA MT radius) at the oxygen site, giving rise to $\sim$ 5.0/f.u 
total magnetic moment, which is again in agreement with the experimental findings.

\begin{figure}
%\centering
\includegraphics[width=6cm]{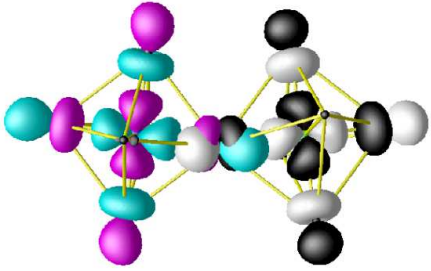}
\caption{\label{fig-2} (Color on-line) Overlap between effective $x^{2}-y^{2}$ orbitals, 
placed at neighboring NiO$_{6}$ and MnO$_{6}$ octahedra of LNMO calculated 
in the monoclinic phase, showing the 
super-exchange path mediated by the corner-shared oxygen. Plotted are the orbital shapes 
(constant-amplitude surfaces) with lobes of opposite signs colored as black(magenta) and
white(cyan) respectively for Mn(Ni).}
\end{figure}

%%{\it Magnetism -}
LNMO being an insulator, the ferromagnetism in this compound is expected
to be dominated by the localized super-exchange kind of interaction, resulting from
the interaction of the half-filled $d$ orbital of one metal ion with the vacant $d$ orbital 
of another metal ion through anion $p$ orbital. 
In the following, we show the feasibility of such a scenario
by considering the calculated hopping interactions between effective
Ni-$d$ orbitals and Mn-$d$ orbitals. The construction of effective Ni- and 
Mn-$d$ orbitals have been carried out with NMTO-
downfolding procedure, by integrating out O and La orbital degrees of 
freedom and keeping active only the Mn- and Ni-$d$ degrees of freedom. 
This procedure generates the effective Ni- and Mn-$d$ orbitals 
(see Fig.\ \ref{fig-2}) which takes into account the renormalization from 
the integrated-out oxygen and also La degrees of freedom. Considering an 
extended Kugel-Khomskii model\cite{12} that includes the hybridization between 
half-filled Ni $e_g$ orbitals and vacant Mn $e_g$ orbitals, the virtual hopping of parallely aligned spins is allowed and is favored over the virtual hopping 
of antiparallely aligned spins due to the energy gain via the Hund's coupling 
$J_H$. The net exchange can be expressed in terms of the sum of the 
square of the Ni $e_g$- Mn $e_g$ hopping interactions, 
$\sum_{m,m'} (t_{e^{m},e^{m'}})^{2}$, the onsite energy differences
$\Delta_{e,e}$, the onsite Coulomb $U$ and the Hund's coupling $J_H$ as:
\begin{equation}
J_{Ni-Mn}^{(1)} = - 4 \frac{\sum_{m,m'} (t_{e^{m},e^{m'}})^{2} J_H}{(U + \Delta_{e,e} - J_{H})(U + \Delta_{e,e})}
\end{equation}
Considering the hoppings, and onsite energy differences for 
Ni-Mn neighbors, $\sum_{m,m'} (t_{e^{m},e^{m'}})^{2}$  turns out to be about 0.2 eV while $\Delta_{e,e}$ turns out
to be about 1.9 eV. Considering a typical $U$ value of 4 eV and $J_H$ of 0.9 eV, 
it gives rise to a value of about -24.5 meV for the FM exchange interaction between Ni and Mn.

However, the Ni-Mn exchange interaction has contribution also from the 
interaction between 
half-filled Ni $e_g$ orbitals and Mn $t_{2g}$ orbitals, which should be 
antiferromagnetic (AFM) in nature, given by:
\begin{equation}
J_{Ni-Mn}^{(2)} = 4 \frac{\sum_{m,m'} (t_{e^{m},t^{m'}})^{2}}{(U + \Delta_{e,t})}
\end{equation}
The summation $m,m'$ in the above runs over half-filled $e_g$ and 
$t_{2g}$ orbitals at Ni and Mn sites, respectively.
The computed sum of squares of the Ni $e_g$- Mn $t_{2g}$ hopping 
interactions in the basis of NMTO-Wannier functions turns out to be about 
0.02 eV, while the on-site energy difference turns out to be about 0.25 eV.
Putting these values in Eqn.(2) it gives rise to a value of about 19 meV 
for $J^{(2)}$. The net exchange interaction, therefore, comes out to be ferromagnetic with a value of about 5 meV\cite{footnote}, 
which is a reasonable estimate considering the rather approximate nature 
of the perturbative approach.
The mean-field T$_c$ computed with this estimate of Ni-Mn exchange 
interaction comes out
to be 350 K compared to experimental estimate of 280 K\cite{note1}.

%%{\it Phonons and dielectric anomaly-}
In the following, we investigate the origin of magnetocapacitance in LNMO manifested
in the dielectric anomaly\cite{1} as a function of magnetic field. The dielectric constant of LNMO is known\cite{1} 
to increase with temperature and exhibit a jump at T$_{jump}$. 
T$_{jump}$ depends sensitively on magnetic field, particularly for small 
fields ($\leq$ 0.1 T). 
Since the electronic contribution to dielectric constant of insulators such
as LNMO is typically much smaller\cite{note2} than the magnitude of jump in
the dielectric constant, we expect the origin of this coupling between 
magnetic field and dielectric response to emerge from the couplings
between spin and structure, {\it i.e.}, phonons.

\begin{figure}[h]
%\centering
\includegraphics[width=9cm]{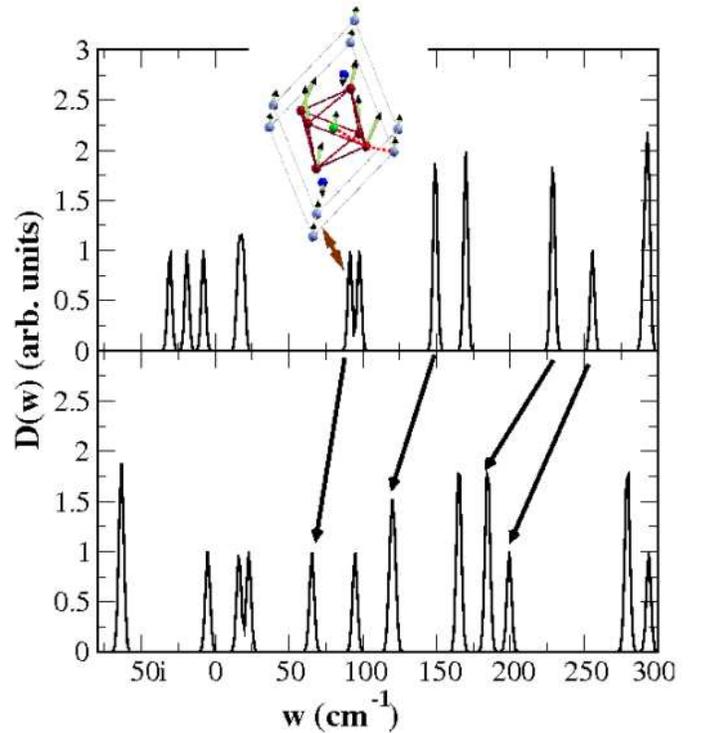}
\caption{\label{fig-3} (Color online) Phonon spectra of rhombohedral LNMO in 
FM (top panel) and FIM (bottom panel) states. The arrows show the shifting of dominant 
IR active phonon modes. The inset shows the displacement of atoms corresponding to the 
lowest frequency IR-active mode. The angle between the red
dotted lines connecting the Ni(at the center of the oxygen-octahedra)-O
and O-Mn (rightmost corner of the cell) is affected by this phonon.}
\end{figure} 

To determine the coupling between spin and various phonons, 
we studied the response of optimized FM rhombohedral structure 
to changes in magnetic ordering: {\it e.g.} changes in phonon
frequencies with changes in magnetic ordering. We determined
changes in phonon frequencies upon changing the neighboring
spin to antiparallel, the so called Ferrimagnetic (FIM) alignment. Note that
any other different spin arrangement other than FM spin arrangement would 
have been equally qualified for this purpose. The charge states of Ni and Mn in the FIM state are found to remain
same as that in the FM state. 
Hellman-Feynman forces acting on atoms in the FIM phase, give us the lowest
order coupling between spins and phonons which is linear in atomic 
displacements. We find that these forces are equal and opposite for pairs of 
atoms, hence this coupling is zero for any IR-active modes and should 
have no direct implications to the observed dielectric anomaly. Next, we 
determined the $\Gamma-$point phonons for the rhombohedral structure with 
FM and FIM ordering. Shifts in the phonon frequencies give
the coupling between spins and atomic displacements at the second order.
Since the rhombohedral structure is unstable at T=0 K, we find two marginally
unstable modes (31$i$ and 19$i$ cm$^{-1}$) which are IR-inactive and couple 
strongly with spins: their frequencies change to 64$i$ and 63$i$ cm$^{-1}$ in
the FIM phase. We find that frequencies of the lowest energy 
IR-active phonons soften from 91.3, 149, 228 and 255  cm$^{-1}$ in the FM 
phase to 65.5, 120, 184 and 199 cm$^{-1}$  respectively (indicated by 
arrows in Fig.\ref{fig-3}), exhibiting a strong coupling with spin. 
This may be compared with the recently studied case
of CdCr$_{2}$S$_{4}$ \cite{rabe} where the significant polar mode was found
only at frequency of 300 cm$^{-1}$. This results in change in static dielectric
constant from 119 in FM state to 221 in the FIM state, which is dominated
by the softest IR-active mode (see inset of Fig.\ref{fig-3}) with a contribution
of 77 and 185, respectively. Atomic displacements in this softest mode, are such
that they would make the Ni-O-Mn angles closer to 180$^{o}$, in an average
sense, leading to enhancement 
of the superexchange interaction. The decrease in phonon frequency for the spin-paired
Ni-Mn in FIM phase compared to spin-antipaired Ni-Mn in FM phase can be explained by 
analyzing the spin Hamiltonian JS$_{i}$.S$_{j}$ and noting that the
magnetic super-exchange coupling J depends on $\angle$ Ni-O-Mn, $\theta$, 
as $cos^{2}(\theta)$. Expanding
$cos^{2}(\theta)$, about the equilibrium value of $\angle$ Ni-O-Mn,
$\theta_{0}$ $\approx$ 157$^{o}$, and assuming $\theta$ = $\theta_{0}$ + $u$,
$u$ being the displacement, the spin-phonon coupling turns out to be positive
for the term linear in $u$ and negative for the term quadratic in $u$. The
latter effectively gives a positive change in phonon frequency due to
the additional negative sign, associated with FM nature of the
magnetic interaction (cf. Eqn.1.). 

%%{\it Conclusion -} 
To conclude, we carried out first principles DFT 
calculations to examine the electronic and magnetic structure of 
La$_2$NiMnO$_6$, in particular, to understand the origin of the dielectric 
anomaly reported recently. We could 
correctly reproduce the ferromagnetic insulating behavior of the compound,
the ferromagnetism being governed by the super-exchange interaction. 
Our study further showed presence of soft IR-active phonon modes 
exhibiting strong coupling with spin,which explains the observed 
dielectric anomaly. The fact 
that the jump in the dielectric constant at H=0 T, occurs 
at a temperature $T_{jump}$ below $T_c$ happens because close to $T_c$ the
magnetic moment is not fully developed due to thermal fluctuation
while at lower temperature the moment gets fully developed and 
makes the effect of coupling to phonon degrees of freedom appreciable 
enough to observe the jump in dielectric constant. This is corroborated
by the fact that $T_{jump}$ becomes closer to $T_c$ upon application of
magnetic field, which helps overcoming the thermal fluctuation
and enhance the magnetization. Note that superexchange driven 
B-site magnetism based dielectric anomaly discussed here is complementary 
to the mechanism of the low-temperature dielectric anomaly discussed for 
EuTiO$_{3}$\cite{eutio}, which is A-site based weak magnetism driven.

{\it Acknowledgment-} 
T.S.-D and D.D.S acknowledge 
support of Swarnajayanti and J.C.Bose grants. U.V.W. thanks SNBose 
Centre for hospitality during the execution of the work and
CCMS at JNCASR for partial support.

%%\section{\label{sec:references}References}


\begin{thebibliography}{99}
\bibitem{8_1} We have also carried out volume optimization of the crystal structure both for RH 
and monoclinic phases. The GGA theoretical lattice parameters are very close to the experimental 
estimates [by less than 0.5$\%$]: 5.537 $\AA$ and 5.478 $\AA$ instead of experimental estimates of 
5.510 $\AA$ and  5.474 $\AA$ for monoclinic and RH phases respectively. 
The structural co-ordinates change at the most by 0.2$\%$ for the monoclinic phase and less than 
0.1$\%$ for the RH phase.
\bibitem{footnote} In this simplictic formulation, we have assumed that the value 
of $U$ and $J_H$ to be the same between Ni and Mn. While $J_H$ is 
fairly constant between different elements, $U$ can vary. Mn is expected to have
a smaller $U$ value than Ni. Considering a $U$ value of 3 eV for Mn and 5 eV for
Ni, one gets an average U value of 4 eV. Changing U by 5 to 10$\%$, 
we find that the calculated effective J turn out to be having a value between 
4 - 7 meV.
\bibitem{note1} We also attempted to extract the net magnetic coupling, $J$, by taking
the total energy difference between FM and FIM configuration of
Ni and Mn spins. LDA total energy differences are however known to overestimate the
value of $J$ due to LDA overbinding problem. Our computed $J$ turned out to be
about 14 meV, which overestimate the perturbatively computed $J$ by a factor of about 
3. Introduction of spin-orbit coupling changes the magnitude of $J$ 
by less than 1$\%$, maintaining the FM nature of $J$.
\bibitem{note2} This is supported by our ab-initio calculation, which gives 
$\epsilon^{\infty}$ equal to 22 and 20 for two different magnetic phases, namely FM and ferrimagnetic phases.
\end{thebibliography}
\end{document}